\documentclass[review]{elsarticle}

\usepackage{amssymb}

\begin{document}

\begin{frontmatter}

\title{Comb-like models for transport along spiny dendrites}
\author[lab1]{Vicen{\c c} M\'{e}ndez}\corref{cor}
\ead{vicenc.mendez@uab.cat}
\author[lab2]{Alexander Iomin}
\ead{iomin@physics.technion.ac.il}

\cortext[cor]{Corresponding author}
\address[lab1]{Grup de F\'{\i}sica Estad\'{\i}stica, Departament
de F\'{\i}sica. Universitat Aut\`onoma de Barcelona. Edifici Cc.
08193 Cerdanyola (Bellaterra) Spain}

\address[lab2]{Department of Physics, Technion, Haifa, 32000, Israel}

\begin{abstract}
We suggest a modification of a comb model to describe 
anomalous transport in spiny dendrites. Geometry of the comb
structure consisting of a one-dimensional backbone and lateral
branches makes it possible to describe anomalous diffusion, where
dynamics inside fingers corresponds to spines, while the backbone
describes diffusion along dendrites. The presented analysis
establishes that the fractional dynamics in spiny dendrites is
controlled by fractal geometry of the comb structure and fractional kinetics
inside the spines. Our results show that the transport along spiny
dendrites is subdiffusive and depends on the density of spines in
agreement with recent experiments.

\end{abstract}

\end{frontmatter}

\section{Introduction}

Dendritic spines are small protrusions from many types of neurons
located on the surface of a neuronal dendrite. They receive most
of the excitatory inputs and their physiological role is still
unclear although most spines are thought to be key elements in
neuronal information processing and plasticity \cite{yusteb}.
Spines are composed of a head ($\sim 1$ $\mu$m) and a thin neck
($\sim 0.1$ $\mu$m) attached to the surface of dendrite (see Fig.
1). The heads of spines have an active membrane, and as a
consequence, they can sustain the propagation of an action
potential with a rate that depends on the spatial density of
spines \cite{pcb}. Decreased spine density can result in cognitive
disorders, such as autism, mental retardation and fragile X
syndrome \cite{nim}. Diffusion over branched smooth dendritic
trees is basically determined by classical diffusion and the mean
square displacement (MSD) along the dendritic axis grows linearly
with time. However, inert particles diffusing along dendrites
enter spines and remain there, trapped inside the spine head and
then escape through a narrow neck to continue their diffusion
along the dendritic axis. Recent experiments together with
numerical simulations have shown that the transport of inert
particles along spiny dendrites of Purkinje and Piramidal cells is
anomalous with an anomalous exponent that depends on the density
of spines \cite{santamaria2006, santamaria2011, zeeuw}. Based on
these results, a fractional Nernst-Planck equation and fractional
cable equation have been proposed for electrodiffusion of ions in
spiny dendrites \cite{henry}. Whereas many studies have been
focused to the coupling between spines and dendrites, they are
either phenomenological cable theories \cite{henry,phen} or
microscopic models for a single spine and parent dendrite
\cite{mic,mic2}. More recently a mesoscopic non-Markovian model
for spines-dendrite interaction and an extension including
reactions in spines and variable residence time have been
developed \cite{prl,book}. These models predict anomalous
diffusion along the dendrite in agreement with the experiments but
are not  able to relate how the anomalous exponent depends on the
density of spines \cite{santamaria2011,zeeuw}. Since these
experiments have been performed with inert particles (i.e., there
are not reaction inside spines or dendrites) we conclude that the
observed anomalous diffusion is due exclusively to the geometric
structure of the spiny dendrite. Recent studies on the transport
of particles inside spiny dendrites indicate the strong relation
between the geometrical structure and anomalous transport
exponents \cite{santamaria2011,bbz,kubota}. Therefore, elaboration
such an analytical model that establishes this relation can be
helpful for further understanding transport properties in spiny
dendrites. The real distribution of spines along the dendrite,
their size and shapes are completely random \cite{nim}, and inside
spines the spine necks act as a transport barrier \cite{mic}. For
these reasons we reasonably assume that the diffusion inside spine
is anomalous. So, we propose in this paper models based on a
comb-like structure that mimic a spiny dendrite; where the
backbone is the dendrite and the fingers (lateral branches) are
the spines. The models predict anomalous transport inside spiny
dendrites, in agreement with the experimental results of Ref.
\cite{santamaria2006}, and also explain the dependence between the
mean square displacement and the density of spines observed in
\cite{santamaria2011}.

\begin{figure}[htbp]
\includegraphics[width=0.9\hsize]{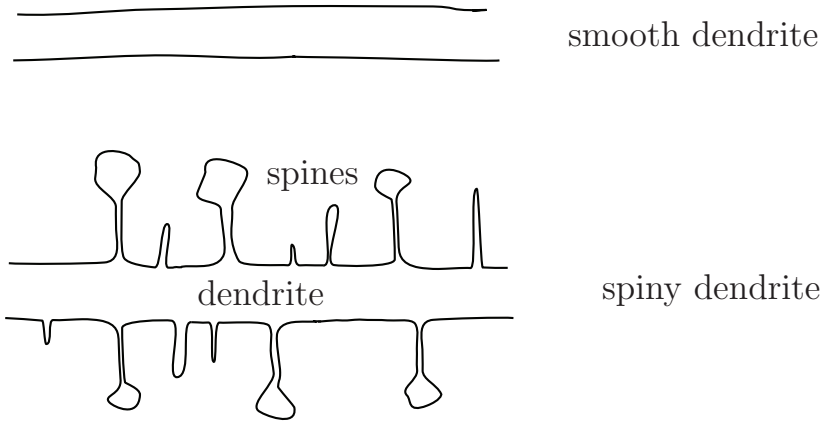}
\caption{Schematic drawing of smooth and spiny dendrites.}
\label{draw}
\end{figure}

\section{Model I: Anomalous diffusion in spines}

Geometry of the comb structure consisting of a one-dimensional
backbone and lateral branches (fingers) \cite{em1} makes it
possible to describe anomalous diffusion in  spiny dendrites
structure in the framework of the comb model. In this case
dynamics inside fingers corresponds to spines, while the backbone
describes diffusion inside dendrites. The comb model is an
analogue of a 1D medium where fractional diffusion has been
observed and explained in the framework of a so-called continuous
time random walk \cite{em1,shlesinger,cmf,klafter}.

Usually, anomalous diffusion on the comb is described by the $2D$
distribution function $P=P(x,y,t)$, and a special behavior is that
the displacement in the $x$--direction is possible only along the
structure backbone ($x$-axis at $y=0$). Therefore, diffusion in
the $x$-direction is highly inhomogeneous. Namely, the diffusion
coefficient is $D_{xx}=D_x\delta(y)$, while the diffusion
coefficient in the $y$--direction (along fingers) is a constant
$D_{yy}=D_y$. Due to this geometrical construction, the flux of
particles along the dendrite is
\begin{equation}\label{Res_1a}
J_{x}=-D_{x}\delta (y)\frac{\partial P}{\partial x}
\end{equation}
and the flux along the finger describes the anomalous trapping
process that occurs inside the spine
\begin{equation}\label{Res_1b}
J_{y}=\left. -D_{y}\frac{\partial ^{1-\gamma }}{\partial
t^{1-\gamma }}\right\vert_{RL}\frac{\partial P}{\partial y}
\end{equation}
where $P(x,y,t)$ is the density of particles and
\begin{equation}\label{der_RL}
\left.\frac{\partial^{1-\gamma}}{\partial t^{1-\gamma
}}\right\vert_{RL}f(t)=\frac{\partial}{\partial t}I_t^{\gamma}f(t)
\end{equation}
is the Riemann-Liouville fractional derivative, where the
fractional integration $I_t^{\gamma}$ is defined by means of the
Laplace transform
\begin{equation}\label{Res_1c}
\hat{\mathcal{L}}\left[I_t^{\gamma}f(t)\right]=
s^{-\gamma}\tilde{f}(s)\, .
\end{equation}
So, inside the spine, the transport process is
anomalous and $\left\langle y^2(t)\right\rangle \sim t^{\gamma }$,
where $\gamma\in(0,1)$. Making use of the continuity equation for
the total number of particles
\begin{equation}\label{Res_1d}
 \frac{\partial P}{\partial t}+\mbox{div}\mathbf{J}=0\, ,
\end{equation}
where $\mathbf{J} = (J_{x},J_{y})$ one has the following evolution
equation for transport along the spiny dendrite
\begin{equation}\label{eb}
\frac{\partial P}{\partial t}-D_{x}\delta (y)\frac{\partial
^{2}P}{\partial x^{2}}-D_{y}\frac{\partial ^{1-\gamma }}{\partial
t^{1-\gamma }}\Big\vert_{RL} \frac{\partial ^{2}P}{\partial
y^{2}}=0.
\end{equation}
The Riemann-Liouville fractional derivative in Eq. (\ref{eb}) is
not convenient for the Laplace transform. To ensure feasibility of
the Laplace transform, which is a strong machinery for treating
fractional equations, one reformulates the problem in a form
suitable for the Laplace transform application.

To shed light on this situation, let us consider a comb in the
$3D$ \cite{bAH}. This model is described by the distribution
function $P_1(x,y,z,t)$ with evolution equation given by the equation
\begin{equation} \label{3d_comb1}
\frac{\partial P_1}{\partial t}-D_{x}\delta (y)\delta(z)
\frac{\partial ^{2}P_1}{\partial x^{2}}-
D_{y}\delta(z)\frac{\partial ^{2}P_1}{\partial y^{2}}-
\frac{\partial ^{2}P_1}{\partial z^{2}}=0\, .
\end{equation}%
It should be stressed that $z$ coordinate is a supplementary,
virtue variable, introduced to described fractional motion in
spines by means of the Markovian process. Thus the true
distribution is $P(x,y,t)=\int_{-\infty}^{\infty} P_1(x,y,z,t)dz$
with corresponding evolution equation
\begin{equation} \label{3d_comb2}
\frac{\partial P}{\partial t}-D_{x}\delta (y) \frac{\partial
^{2}P_1(z=0)}{\partial x^{2}}- D_{y}\frac{\partial
^{2}P_1(z=0)}{\partial y^{2}}=0\, .
\end{equation}
A relation between $P(x,y,t)$ and $P_1(x,y,z=0,t)$ can be
expressed through their Laplace transforms (see derivation in the Appendix)
\begin{equation}\label{3d_comb3}
\tilde{P}_1(x,y,z=0,s)=\frac{\sqrt{s}}{2}\tilde{P}(x,y,s)\, ,
\end{equation}
where $\tilde{P}(x,y,s)=\hat{\mathcal{L}}[P(x,y,t)]$ and
$\tilde{P}_1(x,y,z,s)=\hat{\mathcal{L}}[P_1(x,y,z,t)]$. Therefore, performing
the Laplace transform of Eq. (\ref{3d_comb2}) yields
\begin{equation}
s\tilde{P}(x,y,s)-D_{x}\delta (y) \frac{\partial
^{2}\tilde{P}_1(x,y,z=0,s)}{\partial x^{2}}- D_{y}\frac{\partial
^{2}\tilde{P}_1(x,y,z=0,s)}{\partial y^{2}}=P(x,y,t=0)
\end{equation}
and substituting
relation (\ref{3d_comb3}), dividing by $\sqrt{s}$ and then performing the Laplace
inversion, one obtains the comb model with the fractional time
comb model
\begin{equation} \label{3d_comb4}
\frac{\partial^{\frac{1}{2}}P}{\partial t^{\frac{1}{2}}}-
D_{x}\delta (y)\frac{\partial ^{2}P}{\partial x^{2}}-D_{y}
\frac{\partial ^{2}P}{\partial y^{2}}=0 \, ,
\end{equation}
where $2D_{x,y}\rightarrow D_{x,y}$ and the Caputo
derivative\footnote{To avoid any confusion between the
Riemann-Liouville and the Caputo fractional derivatives, the
former one stands in the text with an index RL: $\frac{\partial
^{\alpha}}{\partial t^{\alpha}}\vert_{RL}$, while the latter
fractional derivative is not indexed $\frac{\partial
^{\alpha}}{\partial t^{\alpha}}$. Note, that it is also convenient
to use Eq. (\ref{LCaputo}) as a definition of the Caputo
fractional derivative.} $\frac{\partial^{\gamma}}{\partial
t^{\gamma}}$ can be defined by the Laplace transform for
$\gamma\in(0,1)$ \cite{mainardi}
\begin{equation}\label{LCaputo}
\hat{\mathcal{L}}\left[\frac{\partial^{\gamma}f}{\partial
t^{\gamma}}\right]=s^{\gamma}\tilde{f}(s)-s^{\gamma-1}f(t=0)\,
.\end{equation} The fractional transport takes place in both the
dendrite $x$ direction and the spines $y$ coordinate. To make
fractional diffusion in dendrite normal, we add the fractional
integration $I_t^{1-\gamma}$ by means of the Laplace transform
(\ref{Res_1c}), as well
$\hat{\mathcal{L}}\left[I_t^{1-\gamma}f(t)\right]=
s^{\gamma-1}\tilde{f}(s)$. This yields Eq. (\ref{3d_comb4}), after
generalization $\frac{1}{2}\rightarrow\gamma\in(0,1)$,
\begin{equation} \label{3d_comb5}
\frac{\partial^{\gamma}P}{\partial t^{\gamma}}-
D_{x}\delta(y)I_t^{1-\gamma} \frac{\partial ^{2}P}{\partial
x^{2}}-D_{y} \frac{\partial ^{2}P}{\partial y^{2}}=0  .
\end{equation}
Performing the Fourier-Laplace transform in (\ref{3d_comb5}) we
get
\begin{equation}
P(k_{x},k_{y},s)=\frac{P(k_{x},k_{y},t=0)-D_{x}k_{x}^{2}P(k_{x},y=0,s)}{%
s+D_{y}k_{y}^{2}s^{1-\gamma }}\, ,  \label{tfl}
\end{equation}%
where the Fourier-Laplace image of the distribution function is
defined by its arguments
$\hat{\mathcal{L}}\hat{\mathcal{F}}_x\hat{\mathcal{F}}_y[P(x,y,t)]=P(k_x,k_y,s)$.
If $P(x,y,t=0)=\delta (x)\delta (y)$, inversion by Fourier over
$y$ gives
\begin{equation}\label{Res_2a}
P(k_{x},y,s)=\frac{1-D_{x}k_{x}^{2}P(k_{x},y=0,s)}{s^{(2-\gamma )/2}\sqrt{%
D_{y}}}\exp \left( -\left\vert y\right\vert s^{\gamma /2}/\sqrt{D_{y}}%
\right) .
\end{equation}
Taking $y=0$ the above equation provides%
\begin{equation}\label{Res_2b}
P(k_{x},y=0,s)=\frac{1}{s^{(2-\gamma )/2}
\sqrt{D_{y}}+D_{x}k_{x}^{2}}
\end{equation}
which inserted into  (\ref{tfl}) yields
\begin{equation}
P(k_{x},k_{y},s)=\frac{1}{s+D_{y}k_{y}^{2}s^{1-\gamma }}\left(
1-\frac{D_{x}k_{x}^{2}}{s^{(2-\gamma )/2}
\sqrt{D_{y}}+D_{x}k_{x}^{2}}\right) . \label{solf}
\end{equation}%
We can calculate the density of particles at a given point $x$ of
the
dendrite at time $t,$ namely $P(x,t),$ by integrating over $y$%
\begin{equation}
P(k_{x},s)=P(k_{x},k_{y}=0,s)=\frac{s^{-\gamma /2}\sqrt{D_{y}}%
}{s^{(2-\gamma )/2}\sqrt{D_{y}}+D_{x}k_{x}^{2}},  \label{Pks}
\end{equation}%
then
\begin{equation} \label{msd_a}
\left\langle x^{2}(s)\right\rangle =-\left. \frac{\partial
^{2}}{\partial
k_{x}^{2}}P(k_{x},s)\right\vert _{k_{x}=0}=\frac{2D_{x}}{\sqrt{D_{y}}}\frac{1%
}{s^{2-\frac{\gamma }{2}}}
\end{equation}
so that
\begin{equation} \label{moment2}
\left\langle x^{2}(t)\right\rangle =\frac{2D_{x}}{\sqrt{D_{y}}}t^{1-\frac{%
\gamma }{2}}.
\end{equation}
Eq. (\ref{moment2}) predicts subdiffusion along the spiny dendrite
which is in agreement with the experimental results reported in
\cite{santamaria2006}. It should be noted that this result is
counterintuitive. Indeed, subdiffusion in spines, or fingers
should lead to the slower subdiffusion in dendrites, or backbone
with the transport exponent less than in usual comb, since these
two processes are strongly correlated. But this correlation is
broken due to the fractional integration $I_t^{1-\gamma}$ in Eq.
(\ref{3d_comb5}). On the other hand, if we invert (\ref{Pks}) by
Fourier-Laplace we obtain the fractional diffusion equation for
$P(x,t)$
\[
\frac{\partial ^{1-\frac{\gamma }{2}}P}{\partial t^{1-\frac{\gamma }{2}}}=%
\frac{D_{x}}{\sqrt{D_{y}}}\frac{\partial ^{2}P}{\partial x^{2}}
\]%
which is equivalent to the generalized Master equation
\begin{equation} \label{ME_a}
\frac{\partial P}{\partial t}=\int_{0}^{t}M(t-t^{\prime
})\frac{\partial ^{2}P(x,t^{\prime })}{\partial x^{2}}dt^{\prime }
\end{equation}%
if the Laplace transform of the memory kernel is given by $
M(s)=\frac{D_{x}}{\sqrt{D_{y}}}s^{\gamma /2}$, which corresponds
to the waiting time PDF in the Laplace space given by
\begin{equation} \label{pdf}
\varphi (s)=\frac{1}{1+\frac{\sqrt{D_{y}}}{D_{x}}s^{1-\frac{\gamma
}{2}}}
\end{equation}
that is $ \varphi (t)\sim t^{-2+\frac{\gamma }{2}}$ as
$t\rightarrow \infty$. The above waiting time PDF is the effective
PDF corresponding to the whole comb and takes into account the
particle  trapping inside spines. Let us employ the notation for a
dynamical exponent $d_w$ used in
\cite{santamaria2006,santamaria2011}. If $d_w=4/(2-\gamma)$ then
the MSD grows as $t^{2/d_w}$. On the other hand, it has been found
in experiments that  $d_w$ increases with the density of spines
$\rho _s$ and the simulations prove that $d_w$ grows linearly with
$\rho _s$. Indeed, the experimental data admits almost any growing
dependence of $d_w$ with $\rho _s$ due to the high variance of the
data (see Fig 5.D in \cite{santamaria2011}). Equation
(\ref{moment2}) also establishes a phenomenological relation
between the second moment and $\rho _s$. When the density spines
is zero then $\gamma=0$, $d_w=2$ and normal diffusion takes place.
If the spine density $\rho _s$ increases, the anomalous exponent
of the PDF (\ref{pdf}) $1-\gamma/2=2/d_w$ must decrease (i.e., the
transport is more subdiffusive due to the increase of $\rho _s$)
so that $d_w$ has to increase as well. So, our model predicts
qualitatively that $d_w$ increases with $\rho _s$, in agreement
with the experimental results in \cite{santamaria2011}.

\section{Model II: L\'evy walks on fractal comb}

In this section we consider a fractal comb model \cite{iom2011} to
take into account the inhomogeneity of the spines distribution.
The comb model is a phenomenological explanation of an
experimental situation, where we introduce a control parameter
that establishes a relation between diffusion along dendrites and
the density of spines. Suggesting more sophisticated relation
between the dynamical exponent and the spine density, we can
reasonably suppose that the fractal dimension, due to the box
counting of the spine necks, is not integer: it is embedded in the
$1D$ space, thus the spine fractal dimension is $\nu\in(0,1)$.
According the fractal geometry (roughly speaking), the most
convenient parameter is the fractal dimension of the spine volume
(mass) $\mu_{\rm spine}(x)\equiv\mu(x) \sim |x|^{\nu}$. Therefore,
following Nigmatulin's idea on a construction of a ``memory
kernel'' on a Cantor set in the Fourier space $|k|^{1-\nu}$
\cite{NIG92} (and further developing in Refs.
\cite{NIG98,Ren,bi2011}), this leads to a convolution integral
between the non-local density of spines and the probability
distribution function $P(x,y,t)$ that can be expressed by means of
the inverse Fourier transform \cite{iom2011}
$\hat{\mathcal{F}}_x^{-1}\left[|k_x|^{1-\nu}P(k_x,y,t)\right]$.
Therefore, the starting mathematical point of the phenomenological
consideration is the fractal comb model
\begin{equation} \label{nu_comb1}
\frac{\partial^{\gamma}P}{\partial t^{\gamma}}-
D_{x}\delta(y)I_t^{1-\gamma} \frac{\partial ^{2}P}{\partial x^{2}}
-D_y\frac{\partial^2}{\partial y^2} \hat{\mathcal{F}}_{k_x}^{-1}
\left[|k_x|^{1-\nu}P(k_x,y,t)\right]=0\, .
\end{equation}
Performing the same analysis in the Fourier-Laplace space,
presented in previous section, then eq. (\ref{Pks}) reads
\begin{equation} \label{nuPks}
P(k_{x},s)=P(k_{x},k_{y}=0,s)=
\frac{s^{-\gamma/2}\sqrt{D_{y}}}{s^{(2-\gamma )/2}
\sqrt{D_{y}}+D_{x}|k_{x}|^{\beta}}\, ,
\end{equation}
where $\beta=3/2+\nu/2$.

Contrary to the previous analysis expression (\ref{msd_a}) does
not work any more, since superlinear motion is involved in the
fractional kinetics. This leads to divergence of the second moment
due to the L\'evy flights. The latter are described by the
distribution $\sim 1/|x|^{1+\beta}$, which is separated from the
waiting time probability distribution $\varphi(t)$. To overcome
this deficiency, we follow the analysis of the L\'evy walks
suggested in \cite{ZKB433,ZK1818}. Therefore, we consider our
exact result in Eq. (\ref{nuPks}) as an approximation obtained
from the joint distribution of the waiting times and the L\'evy
walks. Therefore, a cutoff of the L\'evy flights is expected at
$|x|=t$. This means that a particle moves at a constant velocity
inside dendrites not all times, and this laminar motion is
interrupted by localization inside spines distributed in space by
the power law.

Performing the inverse Laplace transform, we obtain solution in
the form of the Mittag-Leffler function \cite{batmen}
\begin{equation}\label{Pkt}
P(k_x,t)= \mathcal{E}_{1-\gamma/2}
\left(-D|k|^{\beta}t^{1-\gamma/2}\right)\, ,
\end{equation}
where $D=\frac{D_x}{\sqrt{D_y}}$. For the asymptotic behavior
$|k|\rightarrow 0$ the argument of the Mittag-Leffler function can
be small. Note that in the vicinity of the cutoff $|x|=t$ this
corresponds to the large $t$ ($|k|\sim\frac{1}{t}\ll 1$),  Thus we
have \cite{batmen}
\begin{equation}\label{MLasympt}
\mathcal{E}_{1-\gamma/2}
\left(-D|k|^{\beta}t^{1-\gamma/2}\right)\approx \exp\left(
-\frac{D|k|^{\beta}t^{1-\gamma/2}}{\Gamma(2-\gamma/2)}\right)\, .
\end{equation}
Therefore, the inverse Fourier transform yields
\begin{equation} \label{solPxt}
P(x,t)\approx A_{\gamma,\nu}
\frac{Dt^{1-\gamma/2}}{\Gamma(2-\gamma/2)|x|^{(5+\nu)/2}}\, ,
\end{equation}
where $A_{\gamma,\nu}$ is determined from the normalization
condition\footnote{The physical plausibility of estimations
(\ref{MLasympt}) and (\ref{solPxt}) also follows from the
plausible finite result of Eq. (\ref{solPxt}), which is the
normalized distribution $P(x,t)\sim 1/|x|^{(3+\nu+\gamma)/2}$,
where $|x|=t$.}. Now the second moment corresponds to integration
with the cutoff at $x=t$ that yields
\begin{equation} \label{msd_b}
\left\langle x^{2}(t)\right\rangle =
K_{\gamma,\nu}t^{\frac{3-\gamma-\nu}{2}}\, ,
\end{equation}
where
$K_{\gamma,\nu}=\frac{4A_{\gamma,\nu}D_{x}}{(1-\nu)\Gamma(2-\gamma/2)\sqrt{D_{y}}}$
is a generalized diffusion coefficient. Transition to absence of
spines means first transition to normal diffusion in fingers with
$\gamma=1$ and then $\nu=0$ that yields
\begin{equation}\label{END}
\left\langle x^{2}(t)\right\rangle =K_{1,0}t\, .
\end{equation}

\section{Discussion}

The present analysis establishes that the fractional dynamics in
spiny dendrites can be described by two parameters, related to the
fractal geometry of spines $\nu$ and fractional kinetics inside
the spines $\gamma$. Summarizing, the most general
phenomenological description can be performed in the framework of
the fractional Fokker-Planck equation (FFPE)
\begin{equation}\label{FFPE}
\frac{\partial^{\alpha}P}{\partial t^{\alpha}}=K_{\alpha,\beta}
\frac{\partial^{\beta}P}{\partial |x|^{\beta}}\, ,
\end{equation}
where, for the present analysis $\alpha=(2-\gamma)/2$ and
$\beta=(3+\nu)/2$; in general case $\alpha\in (0,1)$ and $\beta\in
(1,2)$ are arbitrary.

For $\beta=2$, we arrive at the first model, presented  in Sec.
II, where we deal with a one temporal control parameter $\gamma$
only. In this case, anomalous transport in dendrites, described by
the dynamical exponent $d_w$, is characterized by anomalous
transport inside spines, described by the transport exponent $\gamma$.
The obtained relation $d_w=\frac{4}{2-\gamma}$ also establishes a
relation between the dynamical exponent and the density of spines
and is in agreement with recent experiments
\cite{santamaria2006,santamaria2011,zeeuw}.

In the second model we suggested a more sophisticated relation
between the dynamical exponent and the spine density. In this case
$\beta=(3+\nu)/2<2$ depends on fractal dimension of spines, and
this leads to an essential restriction for Eq. (\ref{FFPE}). The
first one is a cutoff of the L\'evy flights at $|x|=t$ that leads
to a consequence of laminar and localized motions \cite{ZK1818}
and yields a finite second moment $\left\langle
x^{2}(t)\right\rangle \sim t^{2+\alpha-\beta}$. When $\alpha=1/2$
and $\beta=2$ the FFPE (\ref{FFPE}) corresponds to the continuous
comb model, namely spine dendrites with the maximal density of
spines. For $\alpha=1/2$ and $\beta=3/2$ this model corresponds to
smooth dendrites. Apparently, another physically sound transition
to limiting case is possible for $\nu=1$ and $\gamma=0$ that
corresponds first to the transition to the continuous model, and
then the transition to $\gamma=0$. This physical control of the
parameters ensures an absence of superdiffusion in Eq.
(\ref{FFPE}). Another important question is what happens in
intermediate cases. A challenging question here is what is the
fractal dimension of the spine volume.

We conclude our consideration by presenting the physical reason of
the possible power law distribution of the waiting time PDF
$\varphi(t)$ in Eq. (\ref{pdf}). At this point we paraphrase some
arguments from Ref.~\cite{bAH} with the corresponding adaptation
to the present analysis. Let us consider the escape from a spine
cavity from a potential point of view, where geometrical
parameters of the cavity can be related to a potential $U$. For
example, let us consider  spines with a head of volume $V$
and the cylindrical spine neck of the length $L$ and radius $a$,
and the diffusivity  $D$ \cite{bbz,kubota}. In this case, the
potential is $U=VL/\pi a^2$, which ``keeps'' a particle inside the
cavity, while $D\tau_0$ plays a role of the kinetic energy, or the
``Boltzmann temperature'', where $\tau_0$ is a mean survival time
a particle inside the spine. Therefore, escape probability from
the spine cavity/well is described by the Boltzmann distribution
$\exp(-U/D\tau_0)$. This value is proportional to the inverse
waiting, or survival time
\begin{equation}\label{gimp2}
t\sim\exp\left(\frac{U}{D\tau_0}\right)\, .
\end{equation}
As admitted above, potential $U$ is random and distributed by
the exponential law $P(U)=U_0^{-1}\exp(-U/U_0)$, where $U_0$ is an
averaged geometrical spine characteristic. The probability to find
the waiting time in the interval $(t,t+dt)$ is equal to the
probability to find the trapping potential in the interval
$(U,U+dU)$, namely $\varphi(t)dt=P(U)dU$. Therefore, from Eq.
(\ref{gimp2}) one obtains
\begin{equation}\label{gimp3}
\varphi(t)\sim\frac{1}{t^{1+\alpha}}\, .
\end{equation}
Here $\alpha=\frac{D\tau_0}{U_0}\in (0, 1)$ establishes a relation
between geometry of the dendrite spines and subdiffusion observed
in \cite{santamaria2006,santamaria2011} and support application of
our comb model with anomalous diffusion inside spines, which is a
convenient implement for analytical
exploration of anomalous transport in spiny dendrites in the
framework of the continuous-time-random-walk framework.

\section*{Acknowledgments}
This research has been partially supported by the Generalitat de
Catalunya with the grant  SGR 2009-00164 (VM), by Ministerio de
Econom\'{\i}a y Competitividad with the Grant FIS2012-32334 (VM),
by the Israel Science Foundation ISF (AI), and by the US-Israel
Binational Science Foundation BSF (AI).

\section*{Appendix. Derivation of Eq. (\ref{3d_comb3})}
Eq. (\ref{3d_comb3}) is a relationship between the distributions $P_1(x,y,z=0,t)$ and $P(x,y,t)$ in the Laplace space. Both distributions are related through the expression 
\[ P \left( x, y, t \right) = \int_{- \infty}^{\infty} P_1 \left( x, y, z, t
   \right) d z. \]
If we transform the above equation by Fourier-Laplace we get
\begin{equation}
\tilde{P} \left( k_x, k_y, s \right) = \tilde{P}_1 \left( k_x, k_y, k_z =
   0, s \right). \label{A1}
\end{equation}
Then, Eq. (\ref{3d_comb3}) is nothing but a relation between $\tilde{P}_1 \left( k_x,
k_y, k_z = 0, s \right)$ and $\tilde{P}_1 \left( k_x, k_y, z = 0, s \right)$.
To find $\tilde{P}_1 \left( k_x, k_y, k_z, s \right)$ we transform Eq. (\ref{3d_comb1}) by Fourier-Laplace and after collecting terms we find
\begin{equation}
 \tilde{P}_1 \left( k_x, k_y, k_z, s \right) = \frac{1 - D_x k_x^2 P_1
   \left( k_x, y = 0, z = 0, s \right) - D_y k_y^2 P_1 \left( k_x, k_y, z = 0,
   s \right)}{s + k_z^2} \label{A2}
\end{equation}
where the initial condition has been assumed $P_1 \left( x, y, z, t = 0
\right) = \delta \left( x \right) \delta \left( y \right) \delta \left( z
\right)$ for simplicity.
Setting $k_z = 0$ one gets
\begin{equation}
 \tilde{P}_1 \left( k_x, k_y, k_z = 0, s \right) = \frac{1 - D_x k_x^2 P_1
   \left( k_x, y = 0, z = 0, s \right) - D_y k_y^2 P_1 \left( k_x, k_y, z = 0,
   s \right)}{s}
  \label{A3}  
\end{equation}
Inverting Eq. (\ref{A2}) by Fourier over $k_z$ we obtain
\[ \tilde{P}_1 \left( k_x, k_y, z, s \right) = \frac{1 - D_x k_x^2 P_1 \left(
   k_x, y = 0, z = 0, s \right) - D_y k_y^2 P_1 \left( k_x, k_y, z = 0, s
   \right)}{2 \sqrt{s}} e^{- \sqrt{s} \left| z \right|} \]
and setting $z = 0$
\begin{equation}
 \tilde{P}_1 \left( k_x, k_y, z = 0, s \right) = \frac{1 - D_x k_x^2 P_1
   \left( k_x, y = 0, z = 0, s \right) - D_y k_y^2 P_1 \left( k_x, k_y, z = 0,
   s \right)}{2 \sqrt{s}}
   \label{A4} 
\end{equation}   
Combining (\ref{A3}) and (\ref{A4}) one has
\[ \tilde{P}_1 \left( k_x, k_y, z = 0, s \right) = \frac{\sqrt{s}}{2}
   \tilde{P}_1 \left( k_x, k_y, k_z = 0, s \right) \]
and inverting Fourier over $k_x$ and $k_y$ one finally recovers Eq. (\ref{3d_comb3}).

\end{document}